\documentclass{iauc}
\usepackage{graphicx}

\title[UCDs and M32 twins] 
{UCDs in Fornax and Abell 1689}

\author[S. Mieske et al.]   
{S. Mieske$^1$,
M. Hilker$^1$, \break \and L. Infante$^2$}

\affiliation{$^1$Sternwarte der Universit\"at Bonn, Auf dem H\"ugel 71, 53121 Bonn, Germany\break 
email: smieske@astro.uni-bonn.de, mhilker@astro.uni-bonn.de\\[\affilskip]
$^2$Departamento de Astronom\'\i a y Astrof\'\i sica, P.~Universidad Cat\'olica,
Casilla 104, Santiago 22, Chile\break 
email: linfante@astro.puc.cl}

\pubyear{2005}
\volume{198} 
\pagerange{xx--xx}
\date{?? and in revised form ??}
\jname{Near-field cosmology with dwarf elliptical galaxies}
\editors{H. Jerjen, B. Binggeli}
\begin{document}

\maketitle

\begin{abstract}
We review some properties of ultracompact dwarf galaxies (UCDs) in Fornax and of their brighter counterparts in Abell 1689, among which are two M32 twins, and discuss various possible UCD formation scenarios. We note that it is indispensable to carefully take into account the bright end of the globular cluster luminosity function when estimating the number density of UCDs. It is suggested that the search for more luminous UCDs in dense and rich galaxy clusters is the best way towards establishing UCDs as a new class of galaxies.
\end{abstract}

\firstsection 
\section{Introduction}
\noindent Ultracompact dwarf galaxies (UCDs) have recently been proposed as a new class of galaxies, populating the central regions of the Fornax and Virgo clusters (Hilker et al.~\cite{Hilker99}, Drinkwater et
 al.~\cite{Drinkw00}, Hasegan et al.~\cite{Hasega05}). UCDs are placed between the sequence of globular clusters and
dwarf elliptical galaxies in the fundamental plane of stellar 
systems (Drinkwater et al.~\cite{Drinkw03}). One much 
discussed hypothesis on their origin is that they are remnant nuclei of dwarf galaxies stripped in the potential well of their host cluster (Bekki et al.~\cite{Bekki03}). 
It has also been proposed that UCDs actually are no ``galaxies'' but simply very bright globular clusters (GCs) (Mieske et al.~\cite{Mieske02}), or that they are stellar super clusters created in merger events (Fellhauer \& Kroupa~\cite{Fellha02}). Hasegan et al.~(\cite{Hasega05}) report on the discovery of several UCDs in the Virgo cluster with high M/L ratios and propose the M/L ratio as a criterion to distinguish between overluminous GCs and UCDs.\\
In this contribution, we review the luminosity and colour distribution of UCDs in the Fornax cluster (Sect.~\ref{Fornax}) with respect to the various proposed formation scenarios. Furthermore, we report on the search for UCDs and the subsequent discovery of two M32 twins in the very massive cluster Abell 1689 (Sect.~\ref{A1689}).
\section{UCDs in Fornax}
\begin{figure*}
\begin{center}
\includegraphics[width=6.6cm]{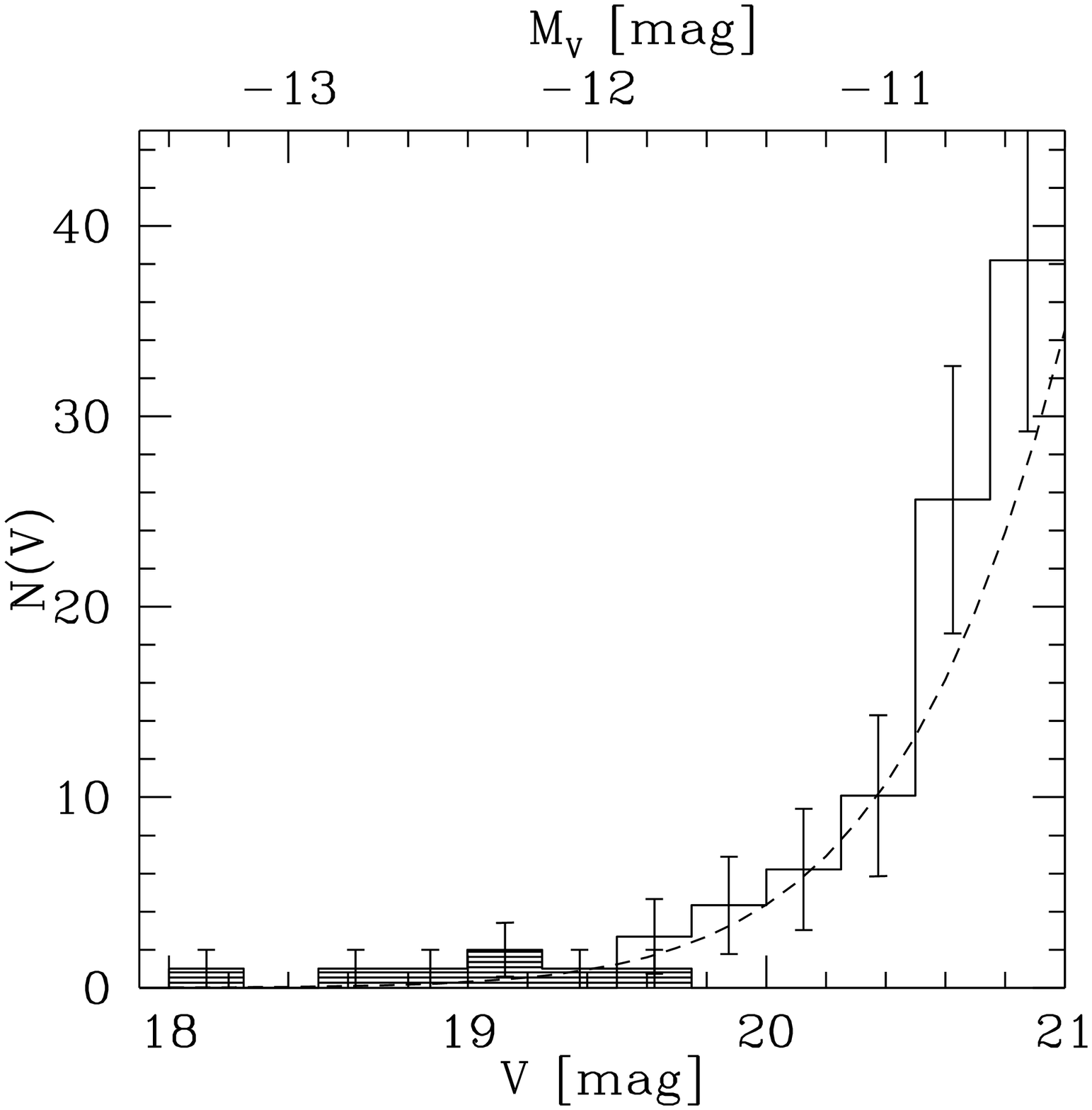}
\includegraphics[width=6.6cm]{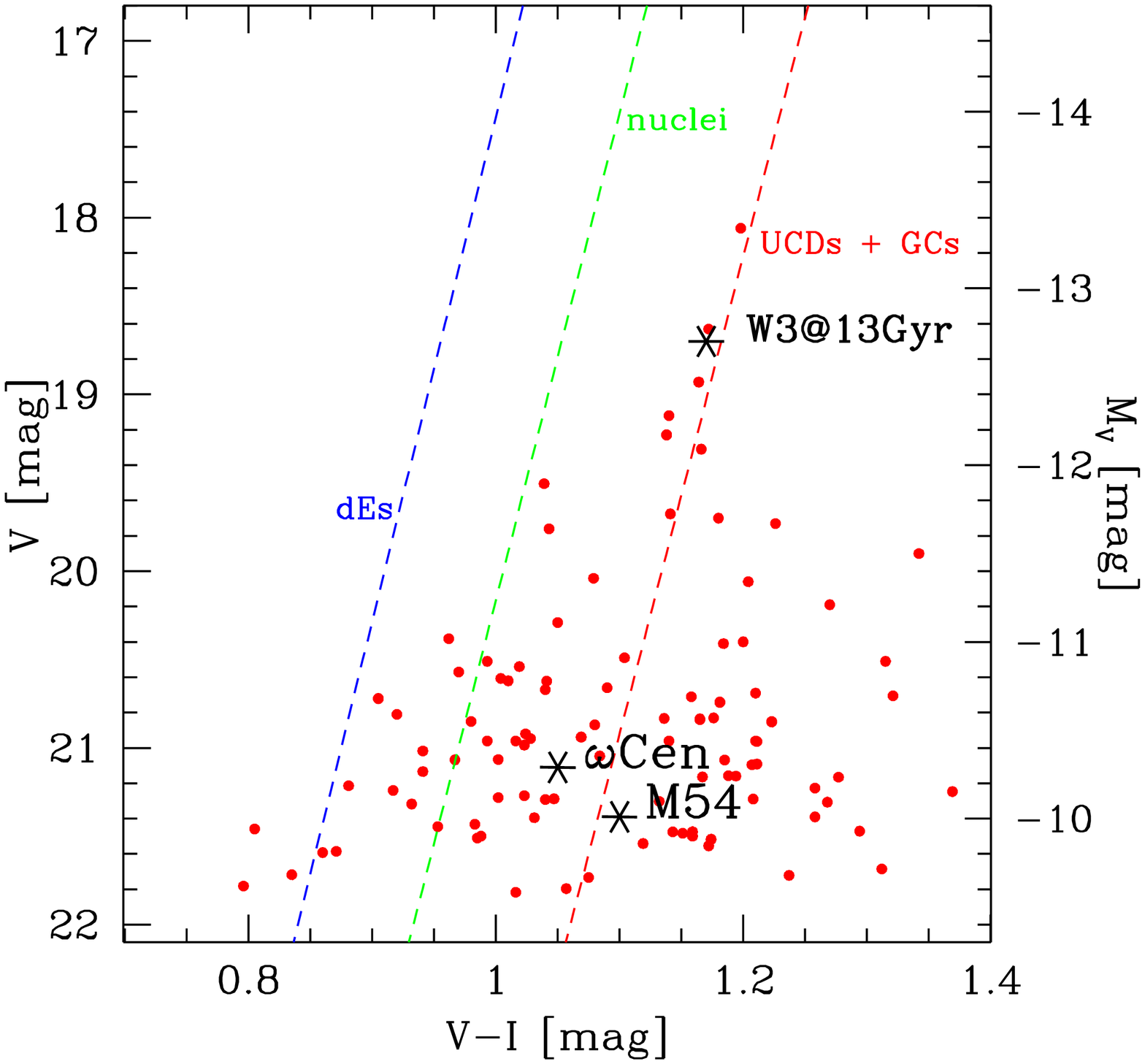}
\end{center}
\caption[]{\label{lf_cmd}{\bf Left:}  {\it Solid histogram:} incompleteness corrected luminosity distribution
$N(V)$ of the compact objects detected in the FCOS (Mieske et al.~\cite{Mieske04a}). {\it Shaded histogram:} luminosity distribution of the UCDs detected by Drinkwater et al.~(\cite{Drinkw03}), photometry taken from the FCOS. {\it Dashed line:} Gaussian LF for the GCLF of NGC 1399, taken
from Dirsch et al. (\cite{Dirsch03}). Note that the Gaussian LF and the luminosity distribution of compact objects are equally scaled and refer to the same total area. {\bf Right:} Colour-magnitude (CM) diagram of the objects in the left panel plus fainter globular clusters with $V>21$ mag from Dirsch et al.~(\cite{Dirsch03}). The asterisks indicate the positions of $\omega$ Cen, M54 -- the two brightest MW globular clusters -- and that of the massive young star cluster W3 (Maraston et al.~\cite{Marast04}) if aged to 13 Gyrs (passive evolution), using BC03 models (Bruzual \& Charlot~\cite{Bruzua03}) and assuming solar metallicity. Dashed lines are from left to right fits to the CM distribution of: all objects plotted except the asterisks; Fornax and Virgo dE,N nuclei (Lotz et al.~\cite{Lotz03}; Fornax dEs (Hilker et al.~\cite{Hilker03}).}
\end{figure*}

\noindent In Fig.~\ref{lf_cmd} we show the distribution of Fornax compact objects in luminosity and colour, as derived in the course of the Fornax Compact Object Survey FCOS (Mieske et al.~\cite{Mieske04b}). We would like to draw the attention of the reader to the following findings:\\
1. The luminosity distribution of compact objects in Fornax with $M_V\ge-11.4$ mag is consistent both in shape and total number of objects with the extrapolated globular cluster luminosity function (GCLF) as determined at fainter magnitudes. That is, {\it it is not necessary to invoke the existence of UCDs with $M_V\ge-11.4$ to explain the luminosity distribution of compact objects. Any survey mapping the luminosity distribution of compact objects fainter than $M_V\simeq -11.4$ mag in Fornax is bound to be totally dominated by the globular cluster population.}\\
2. The UCDs are significantly redder than dE,N nuclei. This may be consistent with the stripping scenario if the star formation in stripped nuclei gets suppressed due to gas removal while being able to continue over more extended periods in the surviving dE,Ns. UCDs plus bright GCs define a colour-magnitude (CM) sequence offset to, but of similar slope as that of dEs. This is consistent with the stripping scenario, because stripped nuclei are expected to trace the colour-magnitude relation of their host galaxies.\\
3. The massive young cluster W3 (Maraston et al.~\cite{Marast04}) matches the colour and luminosity of Fornax UCDs if aged to 13 Gyrs. This is consistent with a scenario in which UCDs are evolved stellar super clusters created in galaxy mergers (Fellhauer \& Kroupa~\cite{Fellha02}).
\label{Fornax}
\section{UCDs in Abell 1689}
\begin{figure*}
\begin{center}
\includegraphics[width=6.5cm]{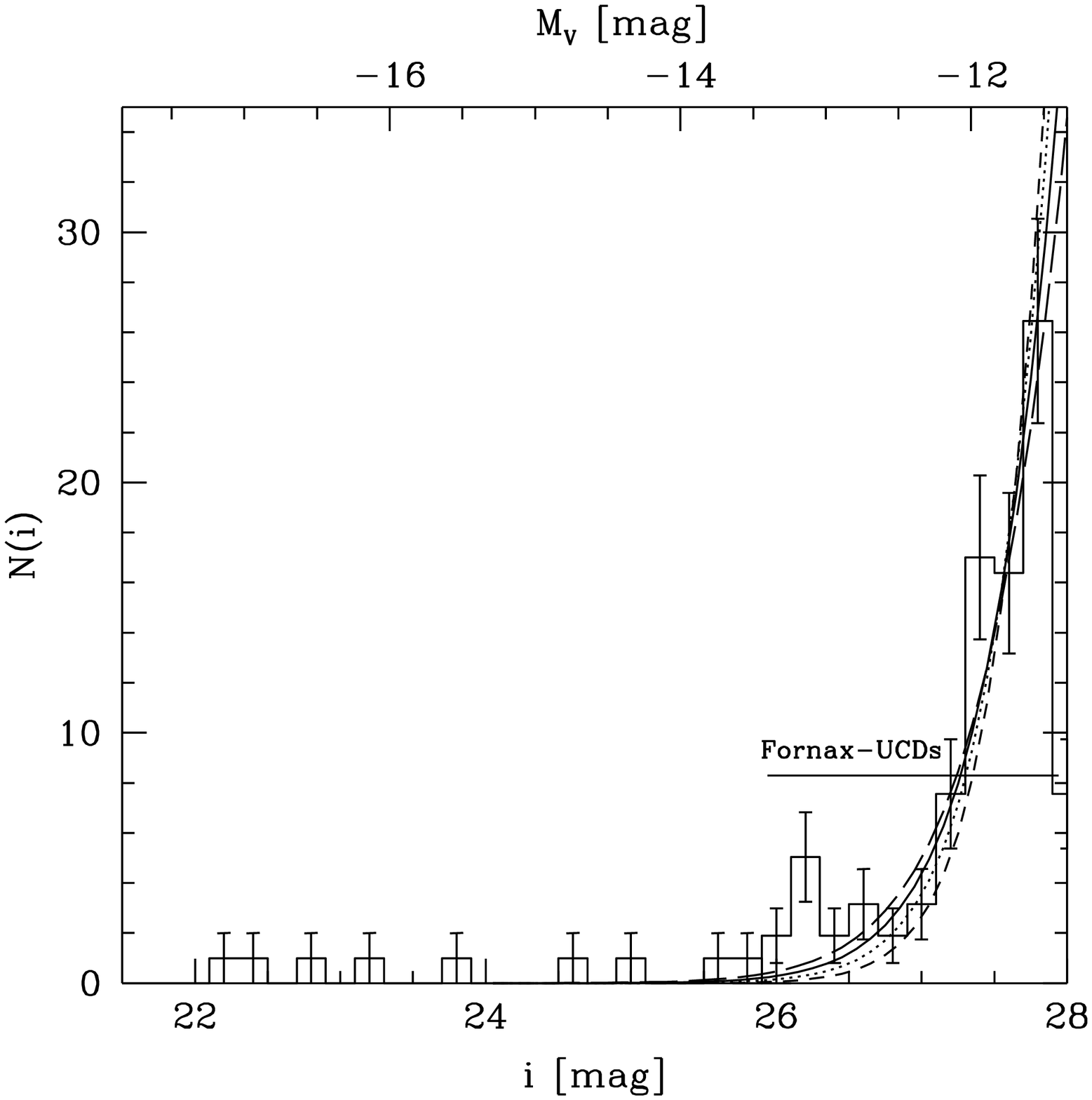}
\includegraphics[width=6.5cm]{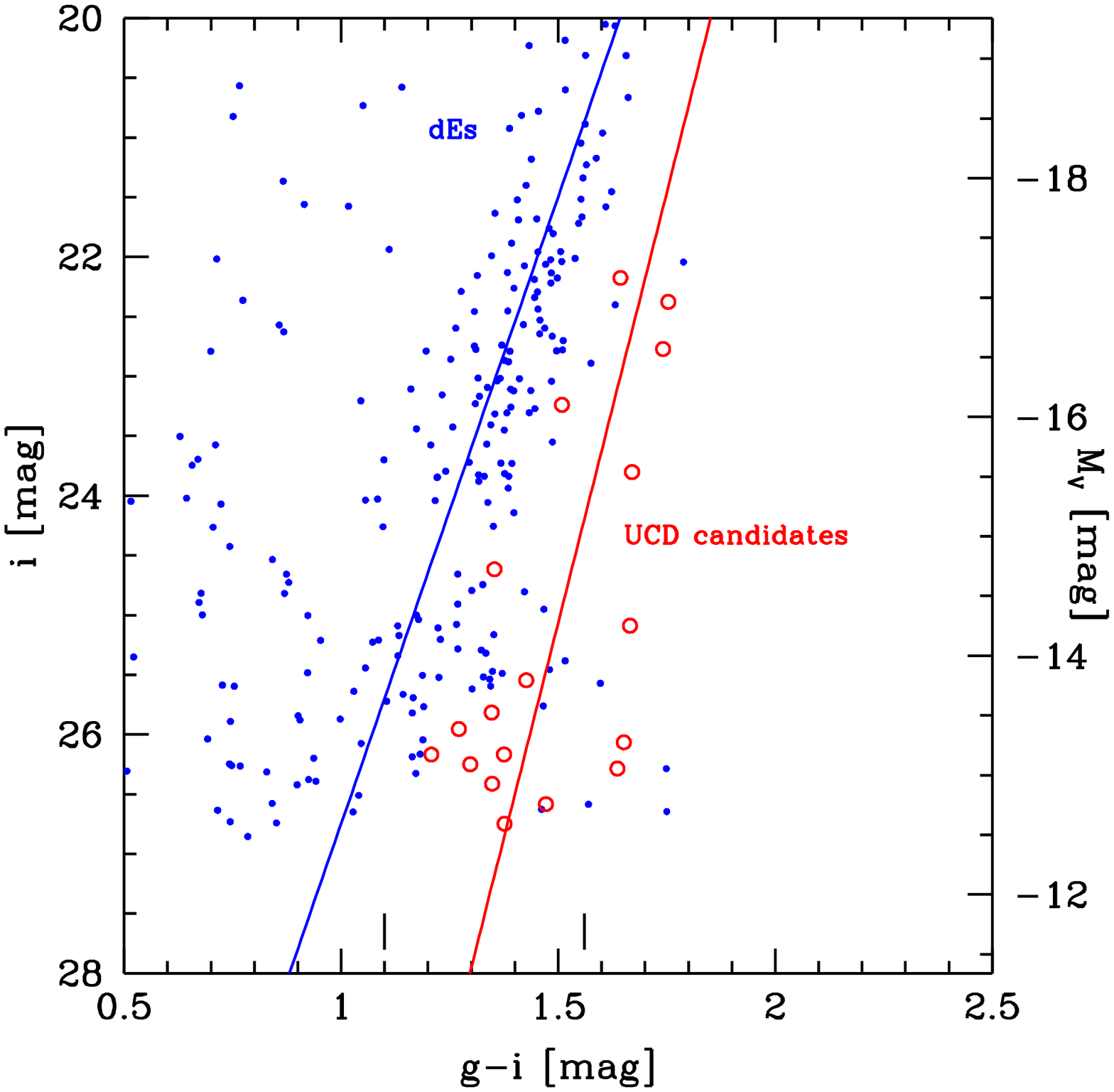}
\includegraphics[width=6.5cm]{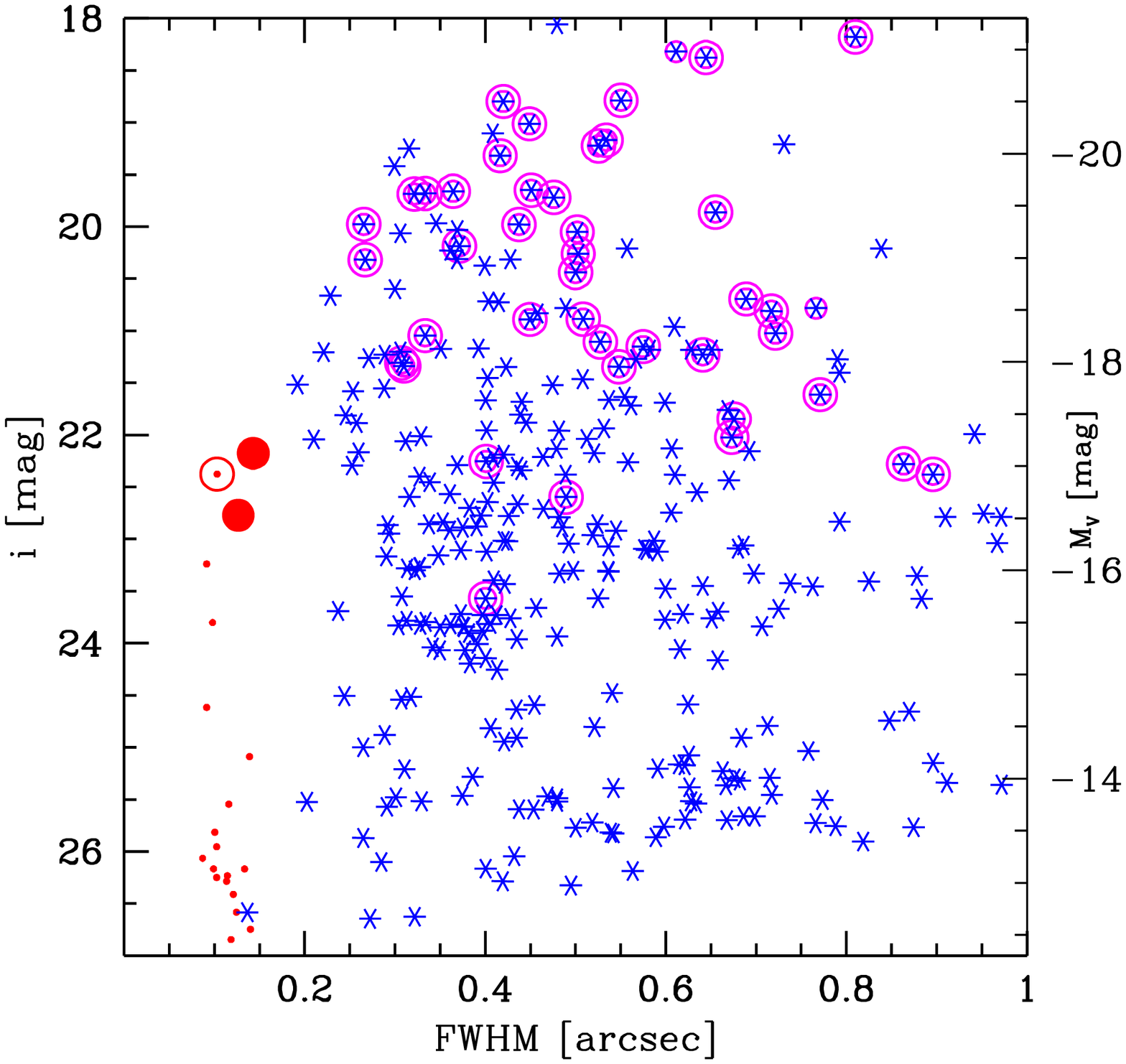}
\includegraphics[width=6.5cm]{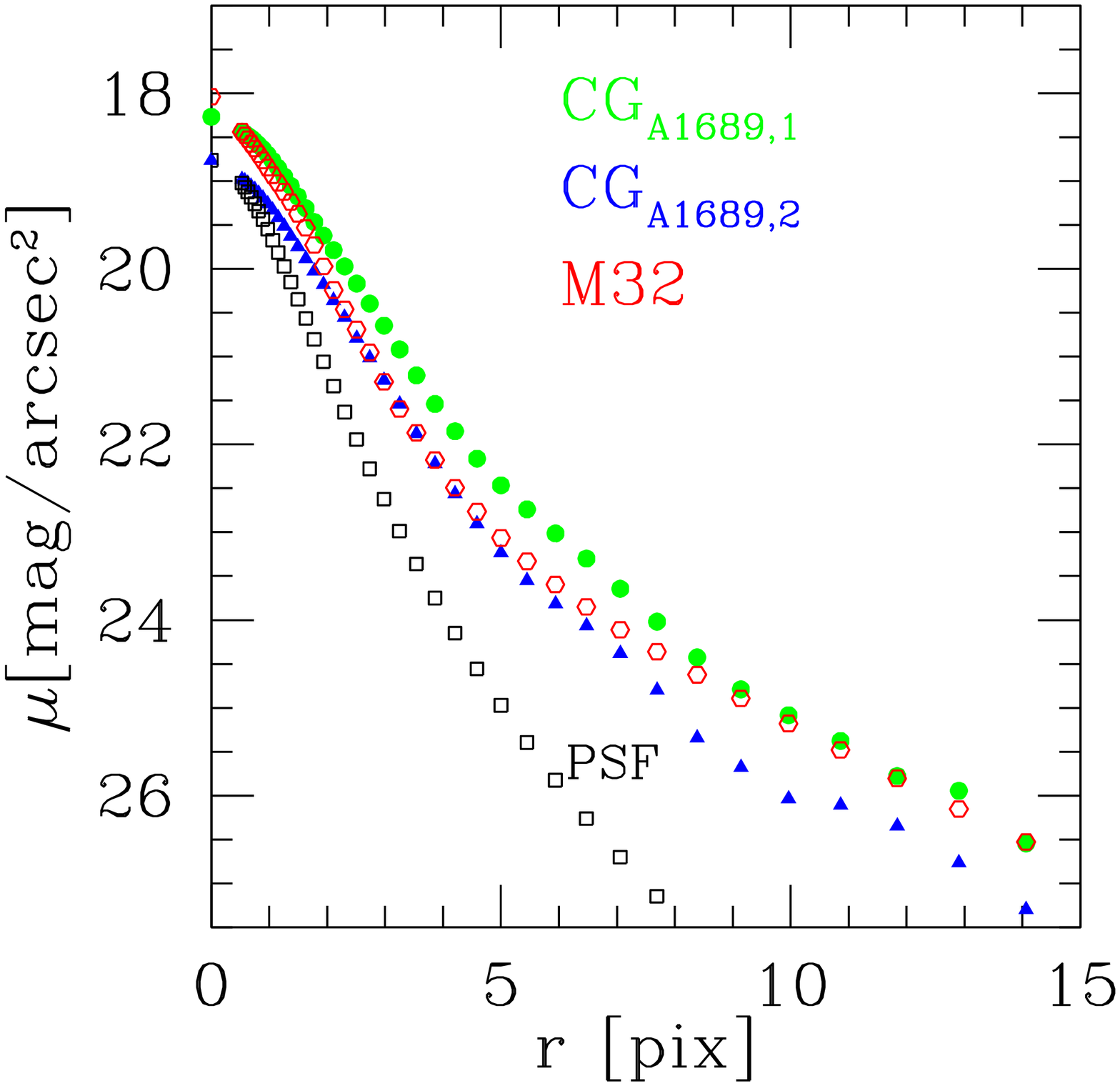}
\end{center}
\caption[]{\label{a1689}{\bf Top left:} Luminosity distribution of UCD candidates in Abell 1689 (Mieske et al.~\cite{Mieske04b}).
The magnitude
range of the UCDs in Fornax is indicated by the horizontal tick.
The long dashed, solid, dotted, short dashed lines
correspond to a Gaussian GCLF at Abell 1689's distance with $\sigma=$ 1.5,
1.4, 1.3, 1.2 mag respectively. {\bf Top right:} CMD in $i$, $(g-i)$ of UCD candidates in A1689 with
$i<27$ mag and $z_{\rm phot}<0.5$ indicated as circles, and resolved objects with
$z_{\rm phot}<0.5$  indicated as
dots (Mieske et al.~\cite{Mieske04b}). The
lines indicate the fitted CM relation to the resolved sources (dashed) and the
UCD candidates (solid). The vertical ticks denote the
expected position of the blue and red peak of the GCLF (Kundu \& Whitmore
\cite{Kundu01}). {\bf Bottom left:} Size-luminosity distribution of normal, i.e. resolved, dwarf galaxies (asterisks) and (unresolved) UCD candidates (dots) in Abell 1689 (Mieske et al.~\cite{Mieske05}). The larger filled circles indicate the two spectroscopically confirmed M32 twins CG$_{\rm A1689,1}$ (at $i=22.2$ mag) and CG$_{\rm
    A1689,2}$ (at $i=22.7$ mag), the UCD candidate with an open circle is a foreground star. All other UCD candidate have not been observed spectroscopically, yet, but observations are underway. The open circles around dwarf galaxy candidates indicate a measured spectroscopic redshift (Czoske~\cite{Czoske04}), double circles are members of Abell 1689. {\bf Bottom right:} Comparison of surface brightness (SB) profiles, with
  scale
0.05$''$/pixel, or 155 pc/pixel at Abell 1689's distance. Filled circles are CG$_{\rm A1689,1}$, filled triangles are CG$_{\rm
    A1689,2}$, open hexagons are M32's profile from 
Graham~(\cite{Graham02}), projected to A1689's distance and PSF convolved. M32 is of intermediate luminosity between CG$_{\rm A1689,1}$ and CG$_{\rm
    A1689,2}$. Squares indicate the PSF profile. }
\end{figure*}

\noindent In Fig.~\ref{a1689} we present the main results of our ACS/VLT search for UCDs in Abell 1689 (Mieske et al.~\cite{Mieske04b} and \cite{Mieske05}). The following findings are obtained:\\
1. The luminosity distribution of UCD candidates with $M_V<-13$ mag cannot be explained by the GCLF of Abell 1689, while for  $M_V>-13$ mag one cannot reject the hypothesis that all UCD candidates are genuine GCs (see also point 1 in Sect.~\ref{Fornax}).\\
2. The colours of UCD candidates in Abell 1689 are redder than normal dwarf galaxies and define a similar slope as the CM relation of normal dwarfs. This is consistent with the stripping scenario (see also point 2 in Sect.~\ref{Fornax}).\\
3. The brighter UCD candidates ($M_V<-13$ mag) form a separate sequence in the size-luminosity plane as compared to normal dwarf galaxies. There is a significant gap in measured size between both populations. This is inconsistent with the hypothesis that UCDs are simply the most compact dEs in a continuous size distribution.\\
4. Two of the three brightest UCD candidates are spectroscopically confirmed as cluster members. According to their total luminosity and surface brightness profiles, they are twins of M32. Finding two of these extremely rare  objects in a massive cluster like Abell 1689 then suggests that tidal forces are responsible for their creation. If the fainter UCD candidates are also confirmed as cluster members by spectroscopy, Abell 1689 would host a continuous sequence of compact objects extending from bright compact ellipticals (cEs) to equivalents of the Fornax UCDs. That is, there might not be a conceptual difference between cEs and UCDs. Both classes of objects might be created by tidal stripping of normal galaxies, leaving only the compact core ``alive''.
\label{A1689}
\section{Conclusions}
\noindent We suggest that in order to estimate the number of bona fide UCDs detected in spectroscopic surveys, one needs to subtract the bright end of the GCLF as extrapolated from fainter magnitudes from the luminosity function of UCD candidates. If no significant overpopulation is detected, it is not possible to confirm the presence of UCDs. Derivation of sizes and masses of UCD candidates (see e.g. Hasegan et al.~\cite{Hasega05}, Drinkwater et al.~\cite{Drinkw03}) seem appropriate ways to distinguish between GCs and UCDs in luminosity regimes where the GCLF has not yet dropped to zero. The existence of a CM relation (see Fig.~\ref{lf_cmd}) may be a criterion to favour a galaxian origin, since no GC system has been found to exhibit such a relation, yet. The search for very luminous and massive UCDs in the densest and most massive galaxy clusters (Mieske et al.~\cite{Mieske04b} and~\cite{Mieske05}) appears the most promising way to definitely establish UCDs as a new class of galaxies.
\begin{acknowledgments}
\noindent We thank the ESO User Support Group for carrying out the DDT
spectroscopy (program 273.B-5008) in service mode. ACS was developed under
NASA contract NAS 5-32864.
\end{acknowledgments}









\end{document}